\documentclass[aps,prl,twocolumn,amsmath,showpacs]{revtex4}
\usepackage{graphicx}

\begin{document}

\title{Observation of anisotropic effect of antiferromagnetic ordering on the
superconducting gap in ErNi$_{2}$B$_{2}$C}

\author{N.~L.~Bobrov \footnote {Email address: bobrov@ilt.kharkov.ua}, V.~N.~Chernobay,
Yu.~G.~Naidyuk, L.~V.~Tyutrina, I.~K.~Yanson }
\affiliation{B.I.~Verkin Institute for Low Temperature Physics and
Engineering, NAS of Ukraine, 47, Lenin Prospect, 61103, Kharkiv,
Ukraine}

\author {D.~G.~Naugle and K.~D.~D.~Rathnayaka}
\affiliation{Department of Physics Texas A\&M University, College
Station TX 77843-4242, USA}

\date{\today}

\begin{abstract}

The point-contact (PC) spectra of the  Andreev reflection $dV/dI$
curves of the superconducting rare-earth nickel borocarbide
ErNi$_{2}$B$_{2}$C (T$_{\rm c}\approx $11~K) have been analyzed in
the ``one-gap'' and ``two-gap'' approximations using the
generalized Blonder-Tinkham-Klapwijk (GBTK) model and the
Beloborod'ko (BB) model allowing for the pair-breaking effect of
magnetic impurities. Experimental and calculated curves have been
compared not only in shape, but in magnitude as well, which
provide more reliable data for determining the temperature
dependence of the energy gap (or superconducting order parameter)
$\Delta $(T). The anisotropic effect of antiferromagnetic ordering
at T$_{N}\approx $6~K on the superconducting gap/order parameter
has been determined: as the temperature is lowered, $\Delta $
decreases by $\sim $25{\%} in the $c$-direction and only by $\sim
$4{\%} in the $ab$-plane. It is found that the pair-breaking
parameter increases in the vicinity of the magnetic transitions,
the increase being more pronounced in the $c$-direction. The
efficiency of the models was tested for providing $\Delta $(T)
data for ErNi$_{2}$B$_{2}$C from Andreev reflection spectra.
\end{abstract}
\maketitle

\section{Introduction}

Quaternary intermetallic nickel borocarbides (hereafter
borocarbides) of the $R$Ni$_{2}$B$_{2}$C type ($R$ is a rare-earth
element) attract special interest (see surveys
\cite{Muller,Muller1,Gupta} and further references) as they
include compounds with rather high superconducting transition
temperatures (up to T$_{\rm c}\approx $17~K, $R$=Lu) and compounds
with different types of magnetic ordering that include states with
commensurate and incommensurate spin-density waves.

Borocarbides have a body centered tetragonal crystalline structure
with the ratio c/a$\sim $3 \cite{Muller, Muller1}. They have a
rather complex Fermi surface (FS) consisting of several sheets
\cite{Bergk, Drechsler}. FS is anisotropic \cite{Muller, Muller1}
and has two characteristic groups of electrons possessing
different Fermi velocities $\nu _{F}$ (this was mentioned in
\cite{Shulga} from the de Haas-van Alphen experiments \cite{Goll,
Nguyen}. The T$_{C}$ of borocarbides is determined not by the
total density of states $N(E_{F})$, but by the contribution to the
density of states which is made by the slow electrons of the nodal
regions \cite{Muller1}. In the normal state the transport
properties of borocarbides, in particular their resistivity
\textit{$\rho $}, are practically isotropic \cite{Muller, Muller1}
because they are related to the groups of electrons that have
relatively high velocities $\nu _{F}$ with lower anisotropy and
are unrelated to the nodal points at the Fermi surface
\cite{Muller1}.

In $R$Ni$_{2}$B$_{2}$C ($R$=Dy, Ho, Er, Tm) compounds the element
$R$ contains 4f-electrons with partially filled f-shells having a
magnetic moment. As a result, T$_{\rm c}$ of these compounds is
appreciably lower in comparison with nonmagnetic borocarbides
($R$=Y, Lu \cite{Muller,Muller1,Gupta}). The object of this study,
ErNi$_{2}$B$_{2}$C, undergoes a superconducting transition at
T$_{\rm c}\sim $11~K \cite{Muller,Muller1,Gupta} and two magnetic
transitions below T$_{\rm c}$, which do not destroy
superconductivity. The AFM ordering occurs at the Neel temperature
T$_{\rm N}\sim $6~K when the Er ions form a transverse-polarized
incommensurate spin-density wave state
\cite{Muller,Muller1,Gupta}. The AFM ordering entails structural
distortions and thus reduces the crystal symmetry from tetragonal
to orthorhombic \cite{Detlefs}. This magneto-elastic effect is
regarded as a structural Jahn-Teller transition \cite{Fil}. The
modulation wave vector of the spin-density waves is practically
independent of temperature and is along the $a$-axis (direction
[100]), or the equivalent $b$-axis (direction [010]), the spins
being aligned along the $b$-axis or the $a$-axis, respectively. As
the temperature lowers further, the compound changes into a weakly
ferromagnetic state with T$_{\tiny{\rm {WFM}}}\sim $2.3~K in which
a spontaneous vortex lattice is formed \cite{Chia}.

To understand the features of the superconducting state in
magnetic borocarbides, it is essential to have information about
the superconducting gap $\Delta $ (magnitude, behavior,
anisotropy, etc.) or the superconducting order parameter (OP). The
investigations  of the gap $\Delta $ in ErNi$_{2}$B$_{2}$C
\cite{Rybalchenko,Watanabe,Yanson,Crespo,Yokoya} give $\Delta
=1.6\div1.82$~meV. Its temperature dependence corresponds on the
whole to the BCS theory. It is noted \cite{Watanabe, Yanson} that
on a paramagnetic -- AFM transition $\Delta $ decreases and the
pair-breaking parameter $\Gamma $ \cite{Dynes} reaches a maximum
in the transition region. The authors \cite{Watanabe} interpreted
the results using the theory \cite{Machida} which predicts a
decrease in $\Delta $ due to spin-density wave gaps that open in
some parts of the FS. On the other hand, the influence of the AFM
transition on $\Delta $ was not observed in subsequent tunnel
measurements on ErNi$_{2}$B$_{2}$C \cite{Crespo}.

Detailed point-contact (PC) spectra of the Andreev reflection in
ErNi$_{2}$B$_{2}$C in two principal crystallographic directions
have been obtained in our recent study \cite{Bobrov}. The analysis
of these spectra shows that:
\begin{enumerate}
\item They are essentially anisotropic and their behavior differs
qualitatively from that in LuNi$_{2}$B$_{2}$C; \item The AFM
ordering lowered the gap; \item The two-gap model can be efficient
at describing the experimental results.
\end{enumerate}
However, the large (six) number of fitting parameters involved in
the two-gap model casts some doubt on the uniqueness of results.
On the other hand, it is quite appropriate to determine to what
extent the one-gap approximation can account for the nontrivial
behavior of the superconducting OP (gap) \cite{Bobrov}. In this
study we have also analyzed the one- and two-gap approximation
within the Beloborod'ko (BB) model \cite{Bobrov,Beloborodko}
allowing for the pair-breaking influence of magnetic ions and on
the basis of the generalized Blonder-Tinkham-Klapwijk (GBTK) model
\cite{Plecenik} allowing for the Dynes pair-breaking parameter
\cite{Dynes}.

\section{Experimental technique}

Here we detail the experimental technique, which was described in
\cite{Bobrov} only briefly because of the space limitations. The
PC measurements were made on ErNi$_{2}$B$_{2}$C single crystals
(T$_{\rm c}\approx $11~K) grown  from the melt (Ames Laboratory,
Prof. P. Canfield's group) and were similar to those used in
\cite{Miao}. The crystals were thin ($0.1\div0.2$~mm) plates with
the $c$-axis normal to the plane of the plate. The surface
pretreatment was similar to that for LuNi$_{2}$B$_{2}$C
\cite{Bobrov1,Bobrov2} (either by etching in a 5{\%} nitric
acid-alcohol solution or by of cleavage). The other electrode was
a high-purity Ag rod or Ag wire of 0.15~mm in diameter. In the
latter case the wire surface was pre-washed in concentrated nitric
acid. The contact was made between the pretreated single crystal
surface and the loop-shaped wire. The use of the wire loop as a
damper improved the mechanical stability of the contacts and made
it possible to measure the characteristics of one of the contacts
(in the $ab$-plane) both on heating from 1.45~K to $T>$11~K
(normal state) and on subsequent cooling from the normal sate to
the starting lowest temperature.

The temperature measurements were made using a continuous flow He
cryostat (its analog is described in \cite{Engel}). An insert was
placed inside the cryostat, which made a PC by touching the sample
surface with the silver electrode at helium temperature. The
typical PC resistance varied from several to tens of Ohms. For
more elaborate investigations, the PCs were selected, which had
the highest possible ``tunneling'' characteristics seen as an
intensive maximum in the $dV/dI(V)$ curve at $V=0$ and the
strongest nonlinearity corresponding to at least a 10{\%} change
in $dV/dI(V)$ in the interval $\pm $8~mV. On some contacts the
$dV/dI(V)$ spectra were measured in the interval from T$_{\rm
{min}}$=1.45~K to temperatures $1\div2$~K higher than T$_{\rm
c}\approx $11~K. The results were quite reproducible irrespective
of the resistance of a particular contact. Therefore, here we
analyze the measurements on two contacts along the principal
crystallographic directions {\-} along the $c$-axis and in the
$ab$-plane. The detailed series of the $dV/dI(V)$ spectra were
obtained at approximately equal temperature intervals. 30 curves
were taken at rising temperature and 34 curves were measured at
lowering temperature in the $ab$-plane, while 49 curves were
measured in the $c$-direction at rising temperature. The contacts
remained stable during the whole period of measurement.

\section{Results and discussion}

\begin{figure}[htbp]
\includegraphics[width=8cm,angle=0]{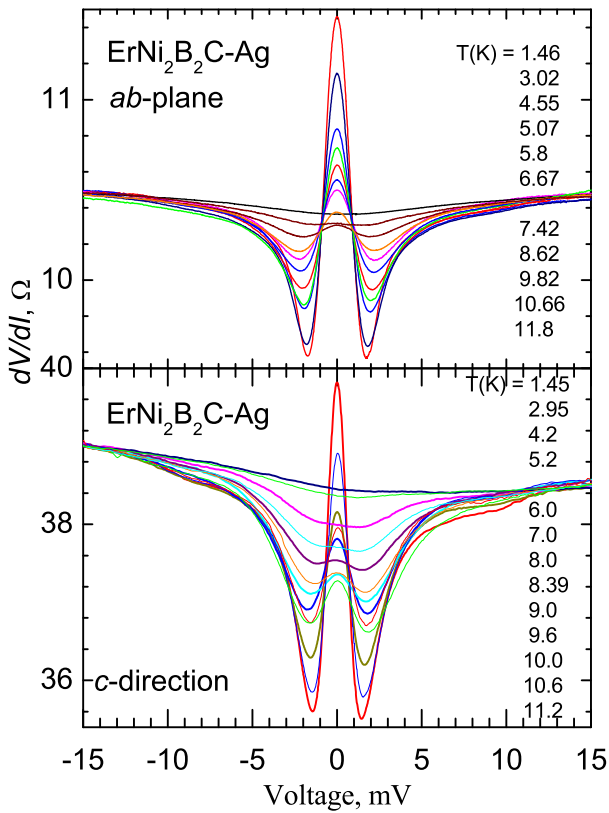}
\caption[]{Differential resistances of ErNi$_{2}$B$_{2}$C-Ag point
contact formed in two principal directions at different
temperatures. To avoid overloading, only a few of the curves are
shown} \label{fig1}
\end{figure}

The temperature series of $dV/dI(V)$ curves measured on two
ErNi$_{2}$B$_{2}C-$Ag contacts in the $ab$-plane and along the $c$-axis are shown in
Fig.\,\ref{fig1}.
Parameters characterizing the contacts and ErNi$_{2}$B$_{2}$C are
described in \cite{Bobrov}. The most important of them is the PC
size (diameter) estimated as $d_{ab}\approx $9.1~nm, $d_{c}\approx
$4.4~nm for the corresponding directions (Fig.\,\ref{fig1}). The
coherence length $\xi $ in this compound increases from 15 to
$\sim $23~nm \cite{Skanthakumar} when $T$ increases from 3~K to
8~K, which satisfies the requirement $d<\xi $ for the theory
\cite{Blonder}.

\begin{figure}[htbp]
\includegraphics[width=8cm,angle=0]{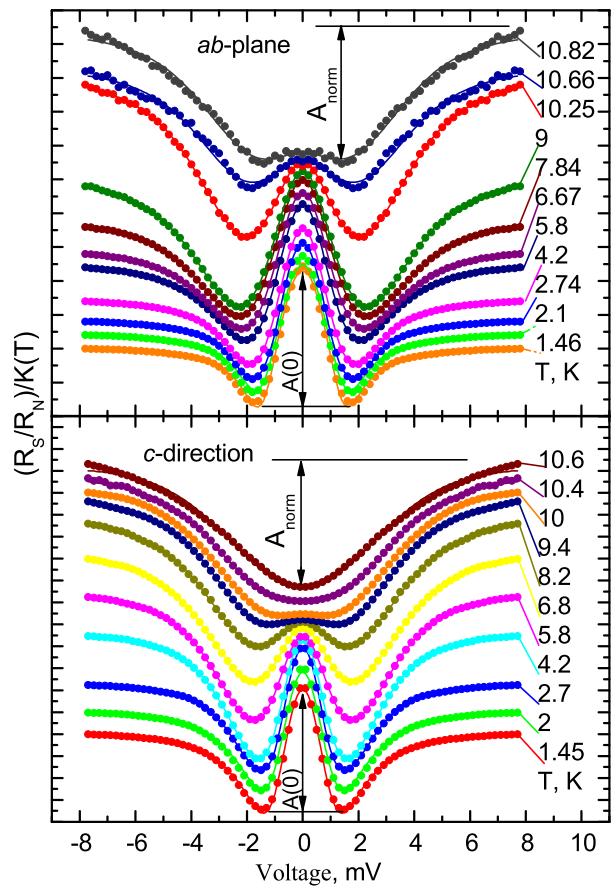}
\caption[]{Symmetrized curves of Fig.\,\ref{fig1} normalized to
the normal state at different temperatures. Points show
experimental data. Lines are theoretical one-gap GBTK calculation.
For visualization, all the curves are reduced to the same
amplitude ($A(0)$=A$_{\rm {norm}}$, see the text)}. \label{fig2}
\end{figure}
\begin{figure}[htbp]
\includegraphics[width=8cm,angle=0]{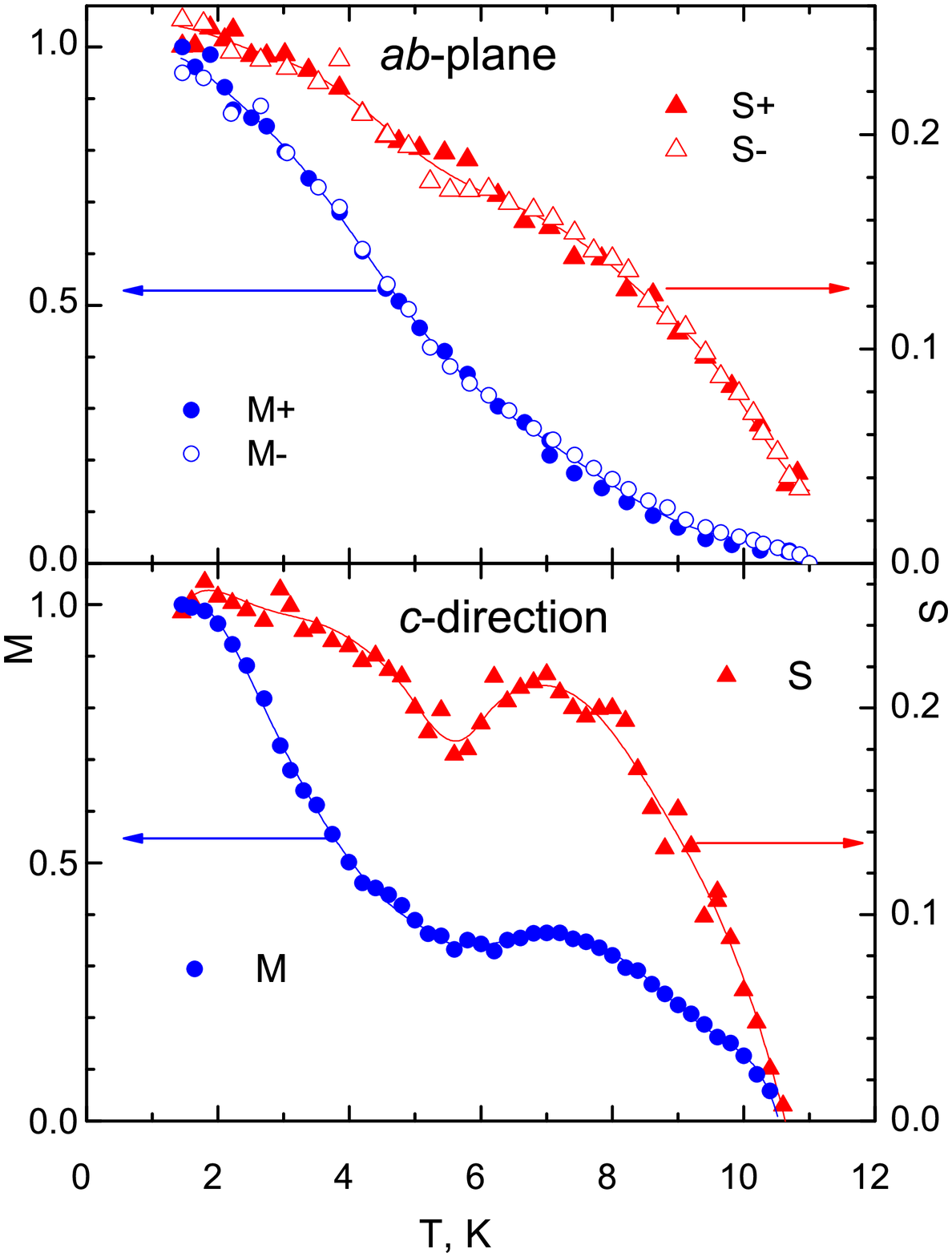}
\caption[]{Temperature dependencies of amplitude coefficients
$M(T)$ obtained from the plots in Fig.\,\ref{fig2} and scaling
factors $S$ calculated in the one-OP BB approximation and
characterizing the intensity of experimental curves in comparison
with theoretical one (see Figs.11,12) in the $ab$-plane and the
$c$-direction. Here and in the subsequent figures the measurements
in the $ab$-plane are marked with solid symbols (rising
temperature,~+) and empty symbols (subsequent cooling,~-~). For
visualization, a polynomial fit is drawn through the points.}
\label{fig3}
\end{figure}

For better visualization, some $dV/dI(V)$ curves of
Fig.\,\ref{fig1} were symmetrized and then scaled by dividing by
$dV/dI(V)$ in the normal state at the lowest temperature and the
bias interval $\pm $8~mV. They are shown in Fig.\,\ref{fig2} along
with the results (lines) of the one-gap calculation within the
GBTK model \cite{Plecenik}. The scaled curved are reduced to equal
amplitude $A_{\rm {norm}}=A(T)/M(T)$. Here $M(T)=A(T)/A(0)$, is a
coefficient, where $A(T)$ is the amplitude at the temperature $T$.
$A(0)$ is the amplitude of the curve normalized to the normal
state at the lowest temperature in the bias interval $\pm $8~mV
($A_{\rm {norm}}=A(0)$). Note that the smooth jump-free dependence
$M(T)$ in the $ab$-plane and $c$-direction (Fig.\,\ref{fig3}) is
indicative of the temperature stability of the point contacts.

\begin{figure}[htbp]
\includegraphics[width=8cm,angle=0]{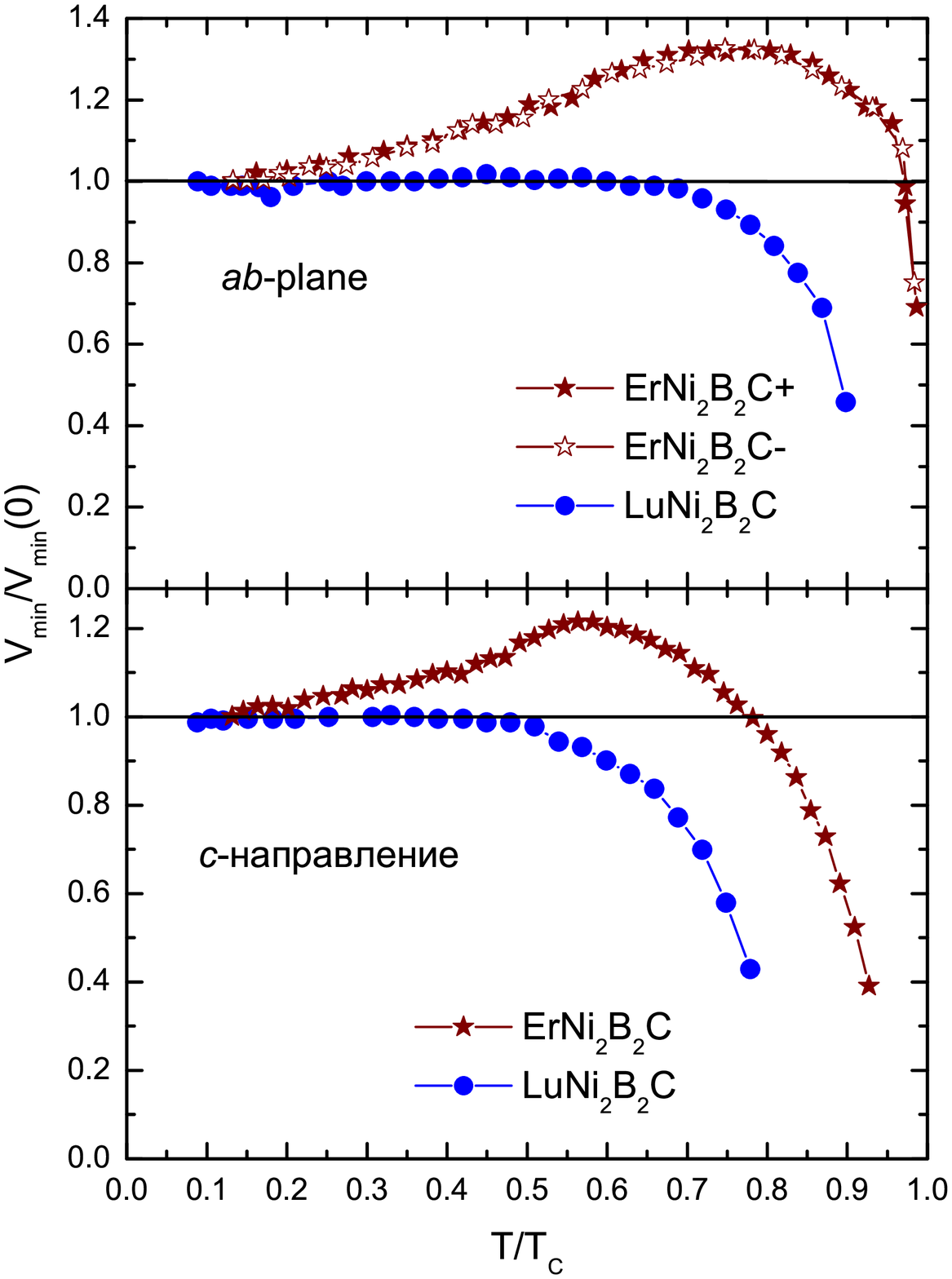}
\caption[]{The temperature dependence of the interminima distance
normalized to the lowest-temperature value for the
ErNi$_{2}$B$_{2}$C-Ag (Fig.\,\ref{fig1}) and LuNi$_{2}$B$_{2}$C-Ag
\cite{Bobrov1} point contacts having similar tunnel parameters MK:
$Z_{Er}^{ab}$ =0.77, $Z_{Lu}^{ab}$ =0.7, $Z_{Er}^{c}$ =0.6,
$Z_{Lu}^{c}$ =0.55.} \label{fig4}
\end{figure}

Of interest is the unusual temperature dependence of the distance
between the minima in $dV/dI$ of the ErNi$_{2}$B$_{2}C-$Ag point
contact (Fig.\,\ref{fig4}).
It differs drastically from the corresponding dependence in
$dV/dI$ of a LuNi$_{2}$B$_{2}C-$Ag point contact having a similar
tunneling parameter Z \cite{Bobrov1, Bobrov2}. It is known that at
low temperatures the half-distance between the minima in the
$dV/dI(V)$ curve of high-tunneling ($Z\sim 1$ S-N point contacts
correlates quite well the superconducting energy gap \footnote{
Even at the lowest temperature the interminima half-distance can
correspond to the OP only when the broadening parameter $\gamma $
\cite{Beloborodko} or $\Gamma $ \cite{Dynes} is zero and the
tunneling parameter Z is non-zero. Besides, the OP can vary in
different regions of the Fermi surface e.q., due to anisotropy.
There is no reason to expect that the interminima half-distance
would coincide with the OP averaged over the Fermi surface.
Therefore, it is expedient to calculate the temperature dependence
of the averaged OP from the fitting results for the theoretical
and experimental curves.}. It should be noted that in the
ErNi$_{2}$B$_{2}C-$Ag PC the two-minima structure of the $dV/dI$
curve persists up to T$_{\rm c}$ in the $ab$-plane and
0.95~T$_{\rm c}$ in the $c$-direction. This is typical of tunnel
contacts (e.g., \cite{Claeson}) and rather unusual in PCs, where
the Andreev reflection is important and the tunnel parameter is
$Z<$1(0.8).

Besides, there is another feature in the ErNi$_{2}$B$_{2}$C spectra that is
unobservable in LuNi$_{2}$B$_{2}$C: the distance between the minima in the
$dV/dI$ curve increases with temperature up to a maximum
slightly above the temperature of the AFM transition. It is reasonable to
attribute this behavior to the magnetic transition in ErNi$_{2}$B$_{2}$C.
Such transitions are absent in LuNi$_{2}$B$_{2}$C.

It is also important that the local critical temperature in the
investigated PCs (at which the main minimum disappears from the
$dV/dI$ curve) practically coincides with T$_{\rm c}$ of the
crystal, which suggests that the properties of the material remain
unaltered in the contact.

As is mentioned in the introduction, experimental data were
analyzed in the one- and two gap approximations using two models:
\begin{enumerate}
\item The traditional GBTK model \cite{Plecenik}, which includes
the broadening parameter $\Gamma $ \cite{Dynes} characterizing
inelastic pair-breaking processes; \item The BB model
\cite{Beloborodko} which introduces the pair-breaking parameter
$\gamma $ to account for the finite lifetime of Cooper pairs due
to the pair-breaking action of (disordered) magnetic moments (in
our case the Er ions possessing a magnetic moment).
\end{enumerate}
The terms "energy gap" (GBTK model \cite{Plecenik}) and "order
parameter" \cite{Beloborodko} used in this model are of equivalent
physical sense (see the detailed discussion in \cite{Beloborodko},
p.014512-3 and are therefore denoted identically with $\Delta $.
This, however, does not refer to the term "energy gap $\Delta
_{0}$" in the BB model \cite{Beloborodko} which differentiates the
superconducting order parameter $\Delta $ and the energy gap
$\Delta _{0 }$ \cite{Beloborodko}. The energy gap $\Delta _{0}$
and the order parameter $\Delta $ are related as
\begin{equation}
\label{eq1}
\Delta _{0} =\Delta \left(1-\gamma ^{2/3} \right)^{3/2}
\end{equation}
Here \textit{$\gamma $}=1/$\tau _{\rm s}\Delta $ is the
pair-breaking parameter, $\tau _{\rm s}$ is the electron mean free
time under spin-flip scattering. When this scattering is absent,
$\tau _{\rm s}$ tends to infinity and the equation describing the
current-voltage characteristics (IVCs) \cite{Beloborodko}
coincides with the corresponding equation of the classical BTK
theory \cite{ Blonder}. The equations describing the IVCs of PCs
within this model are presented in \cite{Beloborodko, Bobrov2}.

\section{One-gap approximation}

The calculation technique of the most popular GBTK model minimizing the
r.m.s. deviations $F$ between the shapes of experimental curves (see \cite{Bobrov2},
Fig.3) faces a certain problem: when the magnitudes of $\Gamma $ and the gap
$\Delta $ become comparable, the error curve for $\Delta $ has no distinct
minimum, which is most typical of the $c$-direction (see Appendix, Fig.18).

Note that in this comparison of the theoretical and experimental
curves the parameter $F$ characterizes only the degree of their
discrepancy in shape, while the distinctions in intensity are
compensated using a scaling factor $S=(dV/dI)_{\rm
{exp}}/(dV/dT)_{\rm {teor}}$. The scaling factor $S$, which
characterizes the intensity ratio between experimental and
theoretical curves, must be equal to 1. Sometimes this requirement
of the GBTK model is violated and we have $S\ne $1. Such factors
and the ways of their selection are considered in Appendix B.

\subsection{Calculation in GBTK and BB models with fixed S}

\begin{figure}[htbp]
\includegraphics[width=8cm,angle=0]{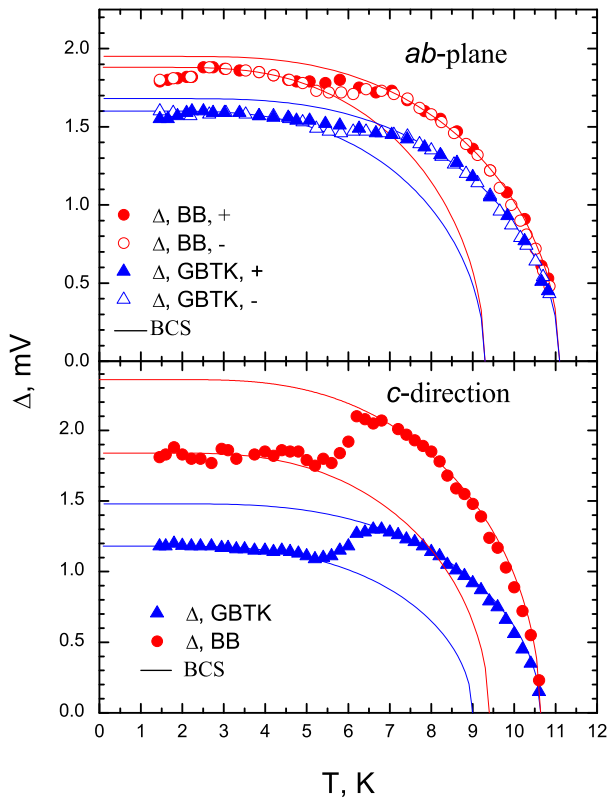}
\caption[]{Temperature dependencies of gaps (GBTK-triangles) and
OP$_{S}$ (BB-circles) calculated for the best-selected scaling
factors $S$ (Fig.\,\ref{fig20}.) for the PC in Fig.\,\ref{fig1}.
Solid lines are BCS extrapolations.} \label{fig5}
\end{figure}

The dependencies $\Delta $(T) in the $ab$-plane and $c$-direction calculated with
properly chosen $S$-factors are shown in Fig.\,\ref{fig5}.

Table 1 contains $\Delta (0)$ estimates (bold type) for both
directions obtained with proper $S$-factors. Also, it includes
GBTK data ($\Delta (0) $) for the maximum ($S$=1) and minimum
($S$=0.25) possible cases. Such $S$ values appear because in this
model the error in $F$ changes only slightly when $S$ deviates
from its proper value (Figs.\,\ref{fig18},\,\ref{fig20}).

Note that in the $ab$-plane the relation $2\Delta/kT_{\rm c}=3.52$
agrees with the BCS theory at $S$=0.31 (our proper selection). In
the $c$-direction the BCS relation $2\Delta/kT_{\rm c}=3.53$
($\Delta $(0)) =1.62~meV in the PM region is achieved at $S$=0.54.
In this case the error in $F$ increases only slightly, see
Fig.\,\ref{fig20}).

\begin{table}[htbp]
\caption[]{Superconducting energy gaps (GBTK) or OP$_{S}$ (BB)
$\Delta $ at different scaling factors $S$ in the $ab$-plane and
$c$-direction.}
\begin{center}
\begin{tabular}{|p{18pt}|p{61pt}|p{20pt}|p{38pt}|p{20pt}|p{6pt}|p{38pt}|}
\hline
\raisebox{-1.50ex}[0cm][0cm]{$S$}&
\raisebox{-1.50ex}[0cm][0cm]{Direction, \par Face}&
\multicolumn{2}{|p{50pt}|}{GBTK} &
\multicolumn{3}{|p{50pt}|}{BB}  \\
\cline{3-7}
 &
 &

$\Delta (0)$ & 2$\Delta /kT_{\rm c}$& $\Delta (0) $ &
\multicolumn{2}{|p{38pt}|}{2$\Delta /kT_{\rm c}$}  \\
\hline
\raisebox{-1.50ex}[0cm][0cm]{0.25}&
$ab$, PM&
1.8&
3.76&
1.95&
\multicolumn{2}{|p{38pt}|}{4.08}  \\
\cline{2-7}
 &
$c$, PM&
2.17&
4.73&
2.36&
\multicolumn{2}{|p{38pt}|}{5.14}  \\
\hline
\textbf{0.31}&
\textbf{$ab$}\textbf{, PM}&
\textbf{1.68}&
\textbf{3.52}&
\multicolumn{3}{ p{50pt} }{}  \\
\cline{1-4}
\textbf{0.65}&
\textbf{$c$}\textbf{, PM}&
\textbf{1.48}&
\textbf{3.23}&
\multicolumn{3}{ p{50pt} }{}  \\
\cline{1-4}
\raisebox{-1.50ex}[0cm][0cm]{1}&
$ab$, PM&
1.03&
2.15&
\multicolumn{3}{ p{50pt} }{}  \\
\cline{2-4}
 &
$c$, PM&
1.23&
2.68&
\multicolumn{3}{ p{50pt} }{}  \\
\cline{1-4}
\end{tabular}
\label{tab1}
\end{center}
\end{table}
The calculation in the BB model gives $2\Delta/kT_{\rm c}=4\div5$.
This correlates with the tunnel investigation on nonmagnetic
YNi$_{2}$B$_{2}$C, in which $2\Delta/kT_{\rm c}=5.2$ for the
maximum gap (see \cite{Nishimori}, Fig.6).

The critical temperature extrapolated to the paramagnetic region
of the BCS curve is close to the values of the bulk compound
T$_{\rm c}$=10.64~K ($c$-direction) and T$_{\rm c}$=11.1~K
($ab$-plane ). At the same time at the AFM-PM transition the
growth of $\Delta $ is essentially dependent on the direction.
According to the BCS extrapolation, $\Delta $ increases by
25-28{\%} in the $c$-direction and only by 4-5{\%} in the
$ab$-plane. Proceeding from the magnetic structure of ground-state
ErNi$_{2}$B$_{2}$C \cite{Muller}, this anisotropic influence of
the magnetic transition on $\Delta $ can be attributed to
orientationally-dependent spin-density waves. The AFM
incommensurate ordering below the Neel temperature induces
spin-density waves whose propagation vector \textbf{q} is in the
$ab$-plane. Such waves reduce the superconducting gap for the
electrons having the wave vector \textbf{k} perpendicular to the
vector \textbf{q}, i.e. for the $c$-direction, due to the
pair-breaking exchange field \cite{Kulic,Brison}. The same
approach was used to interpret the anisotropy of the
superconducting energy gap in the AFM heavy-fermion
URu$_{2}$Si$_{2}$ compound \cite{Naidyuk}. The anisotropic effect
of spin-density waves is also evident in the behavior of the
parameter $M$ (see Fig.\,\ref{fig3}), which has an extremum only
in the $c$-direction. It is important that $M$ characterizes the
"gap minima" intensity of the original $dV/dI$ spectra and is
unrelated to any theoretical model.

The effect of spin fluctuations and the AFM molecular field on the
superconducting gap is determined by the sum rule \cite{Chia1}.
Their competition dictates whether the AFM phase will enhance or
suppress the pair-breaking processes below T$_{\rm N}$. The
temperature dependence of the superconducting gap was calculated
within the Chi-Nagi model \cite{Chi} (Fig.3 in \cite{Chia1}). In
the paramagnetic region the superconducting gap follows the
BCS-dependence and in the AFM region (below T$_{\rm N}$.) its
behavior is determined by the interaction between the
temperature-dependent AFM molecular field and the spin-fluctuation
scattering of conduction electrons at both magnetic rare-earth
ions and nonmagnetic impurities. The molecular field opens AFM
gaps in the some parts of the FS and destroys the superconducting
gaps in them. Nonmagnetic impurities have no effect on the BCS
states of a s-wave superconductor, but they attenuate the AFM
field effect suppressing the pairing states of charge-density or
spin-density waves. The degree of suppression of the
superconducting gap is dependent on the single crystal perfection
-- elastic scattering assists in restoring superconductivity.

\begin{figure}[htbp]
\includegraphics[width=8cm,angle=0]{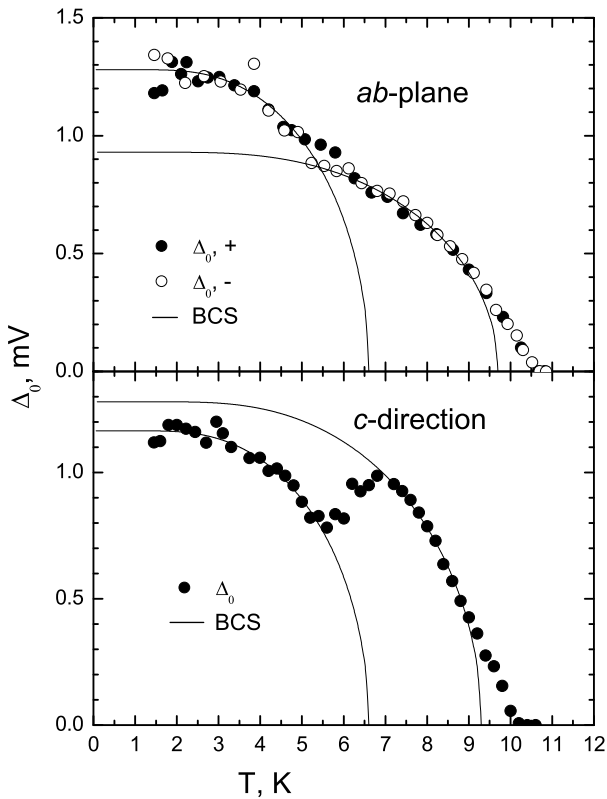}
\caption[]{The temperature dependencies of the BB-calculated gap
$\Delta _{0} =\Delta \left(1-\gamma ^{2/3} \right)^{3/2} $ for the
ErNi$_{2}$B$_{2}$C-Ag PC in the $ab$-plane and the $c$-direction.
Solid lines are BCS extrapolations.} \label{fig6}
\end{figure}

The temperature dependence of the energy gap $\Delta _{0} =\Delta
\left(1-\gamma ^{2/3} \right)^{3/2}$ (Fig.\,\ref{fig6})
can be obtained clearly within the BB model \cite{Beloborodko} as
well. The BCS extrapolations on changing to gapless
superconductivity in the AFM (6.6~K) and PM (9.3-9.6~K) regions
correlate well in both direction with the BCS extrapolations of
T$_{\rm c}$ for the OP$_{S}$ in these regions (Fig.\,\ref{fig5}).
As in the OP case, the gap increases on changing to the PM state
in the $c$-direction and exhibits a monotonic dependence in the
$ab$-plane.

Note that the results of this study and \cite{Bobrov} (decreasing
OPs/gaps in the AFM region) correlate with the temperature
dependencies of the coherence length, the penetration depth and
the critical magnetic fields measured in single crystalline
ErNi$_{2}$B$_{2}$C which has features near T$_{\rm N}$: an
$N$-shaped curve of the coherence length and a local minimum in
the penetration depth (\cite{Gammel}, Fig.1). There is also
indirect evidence for the anisotropy of $\Delta$ which is based on
measurements of the anisotropy of the upper critical magnetic
field H$_{\rm c2}$ (it is known that H$_{\rm c2}\sim \xi ^{-2}\sim
\Delta ^{2})$. A 3-D fourfold modulation of the upper critical
field H$_{\rm c2}$ was measured in the field \textbf{H}$\bot
$\textbf{\textit{c}} at $T$=2~K as a function of the direction in
the $ab$-plane (see \cite{Gammel}, Fig.4, insert). Its shape is
similar to the anisotropic function of the superconducting energy
gap in the model proposed in \cite{Maki}. Note a distinct peak at
T$_{\rm N}$ in the dependence H$_{\rm c2}$(T) in the field along
the $c$-direction (Fig.\,2 in \cite{Gammel}), which indicates
indirectly that the AFM transition reduces the superconducting
gap.
\begin{figure}[htbp]
\includegraphics[width=8cm,angle=0]{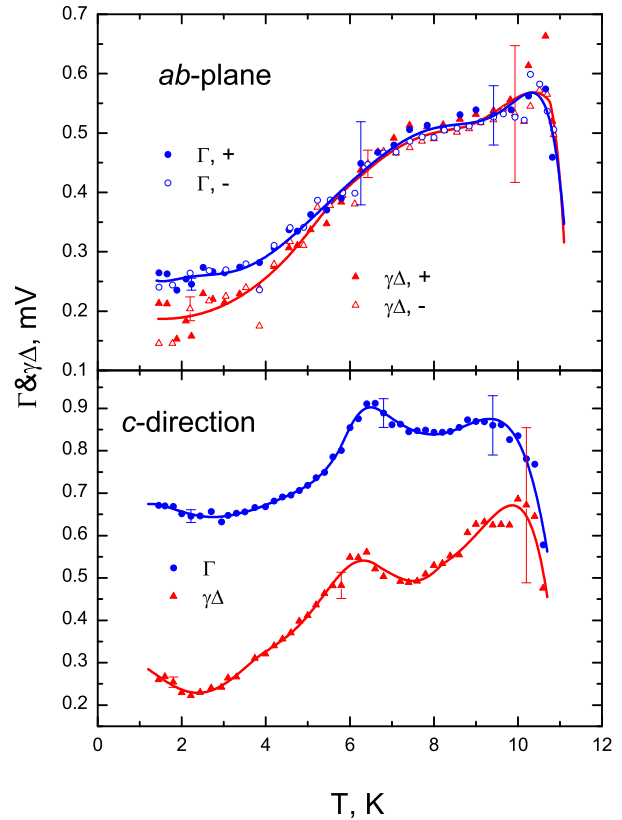}
\caption[]{Temperature dependencies of broadening parameters
$\Gamma$ and pair-breaking parameters  $\gamma$  reduced to the
same units for the ErNi$_{2}$B$_{2}$C-Ag PC in the $ab$-plane and
the $c$-direction. Error bars represent 5{\%} deviations $\Delta$
from the minimum at curve of r.m.s. deviations of the shape of the
theoretical curve from experimental one (see Fig.\,\ref{fig18}).
Since $S$=const was obtained by selecting the broadening
(pair-breaking) parameter, a comparatively slight deviation of the
points from the polynomial fit, the real deviation is much
smaller} \label{fig7}
\end{figure}
Fig.\,\ref{fig7} and Fig.\,\ref{fig8}  illustrate the
pair-breaking parameters $\Gamma $ and $\gamma $ reduced to the
same dimensionality: $\gamma $ is compared with $\Gamma $/$\Delta
$ and $\Gamma $ with $\gamma\cdot\Delta $. The dependencies have
features near the Neel temperature -- a maximum in the
$c$-direction and a smeared shoulder in the $ab$-plane. In the
$c$-direction the pair-breaking parameter increases slightly near
the transition to weak ferromagnetism ($\sim $2~K). In the
$ab$-plane such increase is hard to identify because of the
scatter of experimental points.
\begin{figure*}[htbp]
\includegraphics[width=0.99\linewidth]{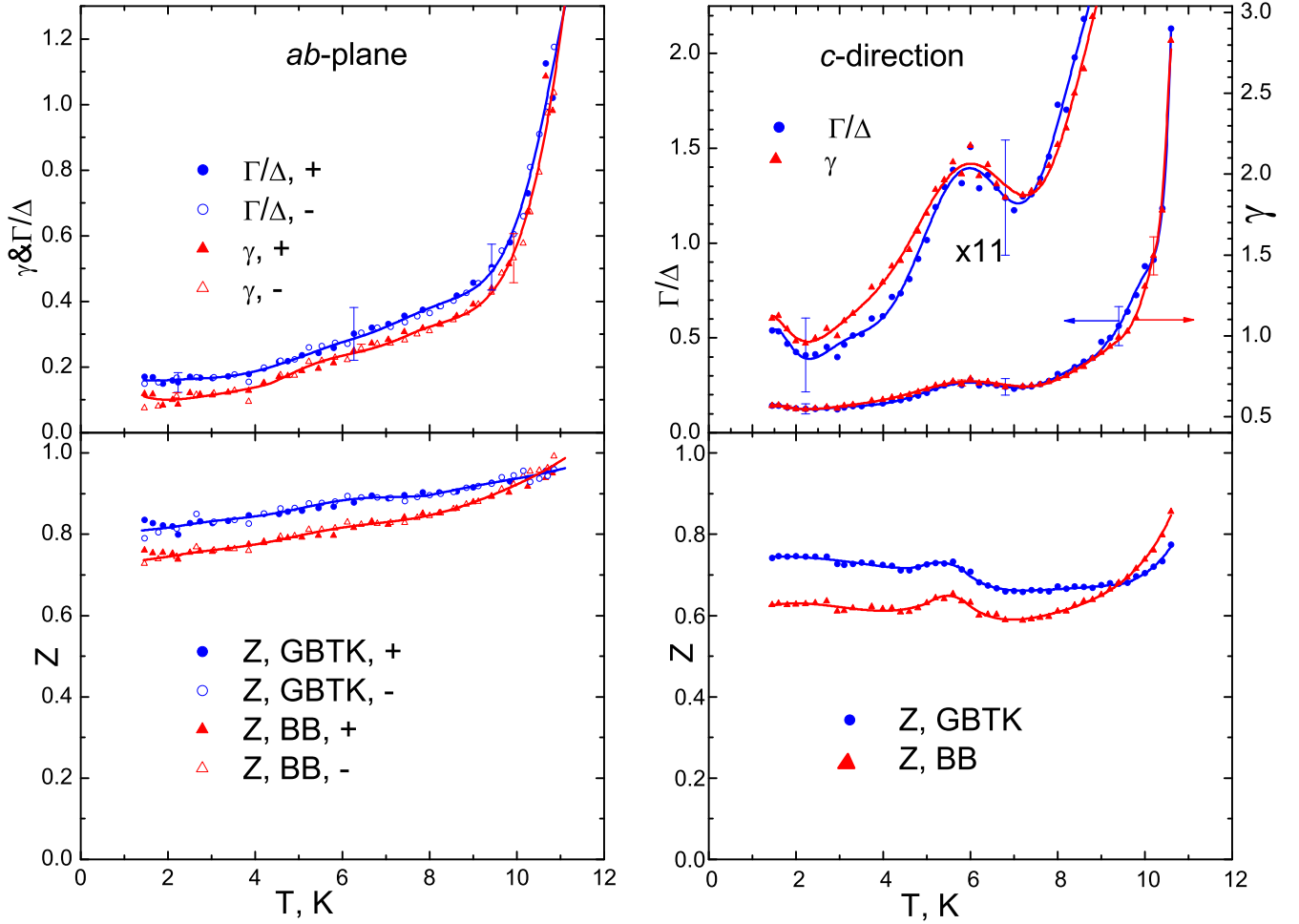}
\caption[]{Temperature dependencies of  the pair-breaking
parameter $\gamma$ and the relative broadening parameter
$\Gamma/\Delta$ (upper pictures) in comparison with the
temperature dependencies of the tunnel parameter $Z$ (lower
pictures) for the ErNi$_{2}$B$_{2}$C-Ag PC in the $ab$-plane and
the $c$-direction. The dependencies $\gamma(T)$ and
$\Gamma/\Delta(T)$ are shown on an enlarged scale in the
$c$-direction to visualize the maximum near the magnetic
transition. The error bars are found as in Figs.\,\ref{fig7}}
\label{fig8}
\end{figure*}
There is a certain correlation between the pair-breaking parameter
$\gamma $ (hence, increasing scattering of superconducting
electrons) and the tunneling parameter $Z$ characterizing the
potential barrier or the scattering at the $N-S$ boundary (see
Fig.\,\ref{fig8}). Thus, the parameter $Z$ may account for both
the elastic scattering intensity and the spin-flip scattering. It
is likely that the growth of these parameters near the AFM
transition is more evident in the $c$-direction because of the
anisotropic influence of the spin-density waves. On approaching
T$_{\rm c}$, $\gamma $ and $Z$ increase in both the $ab$-plane and
the $c$-direction.

\subsection{Calculation with a varying $S$-factor (BB model).}

\begin{figure}[htbp]
\includegraphics[width=8cm,angle=0]{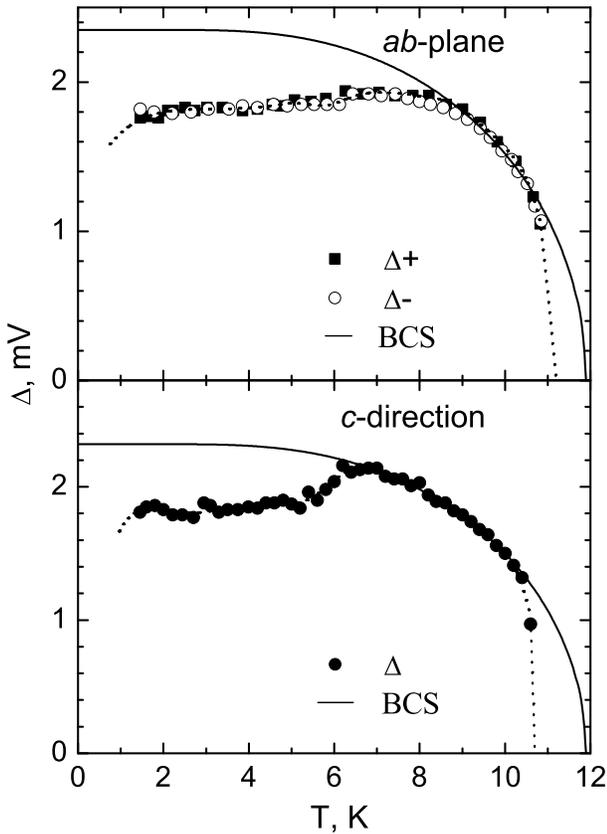}
\caption[]{Temperature dependencies calculated in the one OP
approximation (BB model \cite{Beloborodko}) with $S \neq$~const.
Solid lines are BCS extrapolations} \label{fig9}
\end{figure}

Note that the use of a fixed $S$ in the BB model reduced the
quality of fitting of theoretical to experimental curves. The
reasons for such reduction in quality are considered in Appendix
B. The temperature dependencies of OPs having freely varying $S$
factors are shown in Fig.\,\ref{fig9} for the $ab$-plane and
$c$-direction (two-gap approximation with free $S$ was used in
\cite{Bobrov}).
In both directions the OPs deviate from the BCS curve on
approaching T$_{\rm c}$ (at $T\sim $10~K) and turn zero at T$_{\rm
c}$ of the contact. The critical temperature obtained for these
parameters through a BCS extrapolation is T$_{\rm c}$=11.9~K in
both directions. The temperature dependencies of the scaling
factors $S$ (Fig.\,\ref{fig3}) correlate in shape with those of
the amplitude coefficients $M(T)$. Such temperature dependencies
are due to the density of states curve in this model which differs
from the BCS dependence (see the Appendix), as well as to the
two-gap character of superconductivity in this compound whose OPs
have significantly different magnitudes in the paramagnetic region
(detailed in the following section).

\begin{figure}[htbp]
\includegraphics[width=8cm,angle=0]{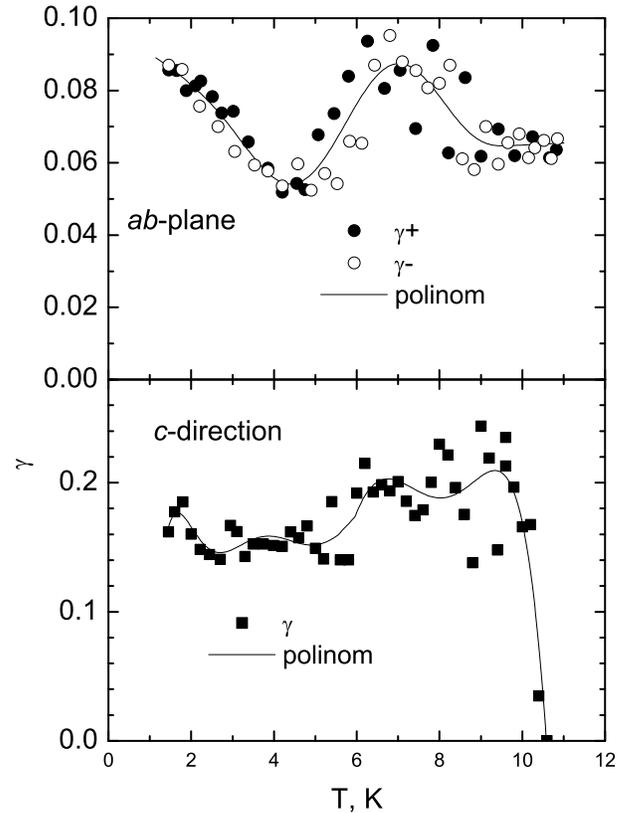}
\caption[]{Temperature dependencies of pair-breaking parameters $\gamma$ in the one
OP approximation with $S\neq$const}
\label{fig10}
\end{figure}

Finally, Fig.\,\ref{fig10} illustrates the temperature
dependencies of the pair-breaking parameter $\gamma $.
On the whole, they are similar to the temperature dependencies of
the pair-breaking parameters for a large OP in the two-gap
approximation \cite{Bobrov} and differ considerably from the
corresponding dependencies in the one-gap approximation obtained
with a fixed scaling factor (Fig.\,\ref{fig8}), which is
particularly evident in the high-temperature region. This may be
because in the latter case we try to hold the scaling factor
invariable at the expense of a certain departure from coincidence
of the shapes of experimental and theoretical curves. In the
vicinity of magnetic transitions pair-breaking parameters also
increase in both $ab$-plane and $c$-direction.

Thus, the behavior of OPs and other parameters estimated in the
one-gap approximation within the BB model with a free factor $S$
is similar qualitatively to the results obtained in the two-gap
approximation \cite{Bobrov}.

\section{Two-gap approximation}

The two-gap approximation assumes that the total conductivity is a
superposition of conductivities from two region (bands) of the FS with
corresponding gaps. This can be expressed for $dV/dI$ as
\begin{equation}
\label{eq2}
\frac{dV}{dI}=\frac{S}{\frac{dI}{dV}\left( \Delta _{1},\gamma
_{1},Z\right) K+\frac{dI}{dV}\left( \Delta _{2},\gamma
_{2},Z\right) \left( 1-K\right) }
\end{equation}
Here the coefficient $K$ accounts for the contribution to the
conductivity from the FS region with a smaller gap $\Delta_{1}$,
$Z$ is the tunnel parameter, $S$ is the scaling factor
characterizing the intensity ratio of the experimental and
theoretical curves, like in the one-gap approximation. This
expression was used to fit experimental curves and to derive the
parameters $\Delta _{1,2}$, $\Gamma _{1,2}$ (or $\gamma _{1,2})$,
$Z$, $S$ and $K$. The calculation technique is detailed in
\cite{Bobrov2}, Appendix.

The average gap is found from the expression
\begin{equation}
\label{eq3} \Delta_{\rm {aver}}=\Delta _{1}K+\Delta _{2}(1-K)
\end{equation}

\subsection{GBTK model}

The temperature dependencies of the larger, smaller and the average gaps
obtained in the two-gap approximation within the GBTK model with a fixed
contribution $K$ are shown in Fig.\,\ref{fig11} (also see Appendix B).
\begin{figure}[htbp]
\includegraphics[width=8cm,angle=0]{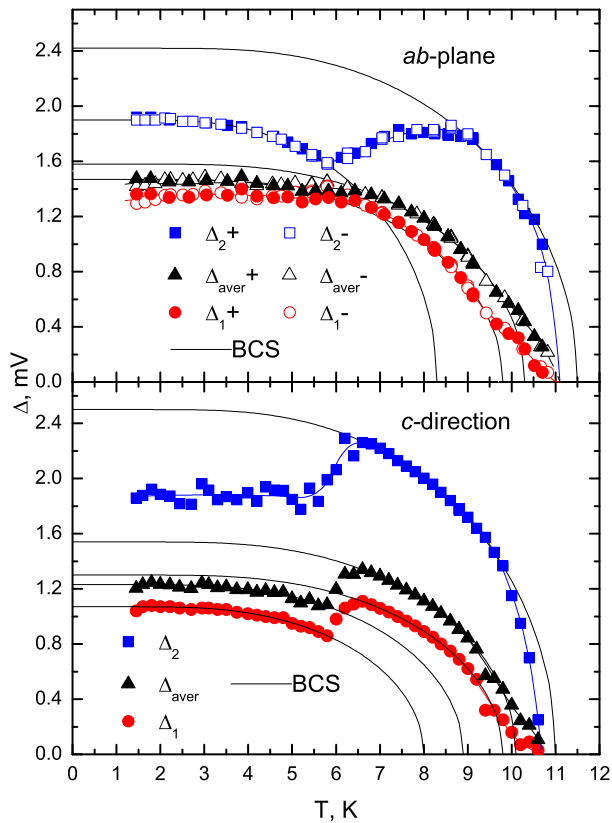}
\caption[]{Temperature dependencies of gaps $\Delta$ (GBTK
\cite{Plecenik} [21]) calculated in the two-gap approximation by
Eq.(\ref{eq2}) for the PC in Fig.\,\ref{fig1}. The average gaps
are $\Delta_{\rm {aver}}=\Delta _{1}K+\Delta _{2}(1-K)$. The
polynomial fit runs through the points outside the BCS
approximations. The scaling factors are $S_{ab}$=0.35 and
$S_{c}$=0.5. The contribution to conductivity from the smaller gap
is $K$=0.8 ($K$=const)} \label{fig11}
\end{figure}
Although $K$ is fixed, the critical temperatures obtained  through
a BCS extrapolation are different for the larger and smaller gaps
in the PM region, which may point to the two-gap character of
superconductivity with a weak interband scattering. Such behavior
is impossible in the case of ordinary OP anisotropy where these is
only one critical temperature, according to Pokrovsky's theorem
\cite{Pokrovsky,Pokrovsky1}. Besides, BCS extrapolation can be
employed to estimate T$_{\rm c}$ in the AFM region for the larger
gap in the $ab$-plane and for the smaller one in the
$c$-direction. The temperature dependencies of the broadening
parameters $\Gamma $ are illustrated in Fig.\,\ref{fig12}.
\begin{figure}[htbp]
\includegraphics[width=8cm,angle=0]{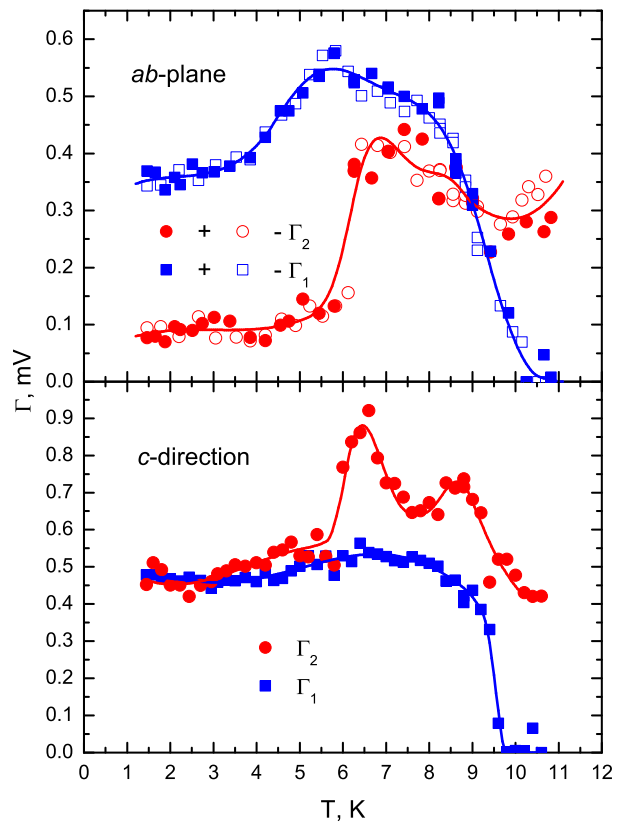}
\caption[]{Temperature dependencies of broadening parameters
$\Gamma$ calculated in the two-gap GBTK model. $\Gamma _{2}$ and
$\Gamma _{1}$ are for the larger ($\Delta _{2}$) and smaller
($\Delta _{1}$) gaps, respectively (Fig.\,\ref{fig11}). For
clarity, a polynomial fit is drawn through the points}
\label{fig12}
\end{figure}
In the AFM region $\Gamma _{1}>\Gamma _{2}$ in the $ab$-plane and
$\Gamma _{1}\sim \Gamma _{2 }$ in the $c$-direction. The
comparison with the one-gap calculation (Fig.\,\ref{fig7}) shows
that the shapes of the curves in Fig.\,\ref{fig7} are closer to
$\Gamma _{2}$, i.e. the smearing of the high-energy part of the
gap is important in the one-gap approximation. The temperature
dependencies of gaps and broadening parameters discussed in this
section are of illustrative character because of $K$=const. The
goal was to show that they have different shapes and T$_{\rm c}$
differs from BCS-extrapolated critical temperatures. Note that the
average gaps shown in this figure practically coincide with those
calculated in the one-gap approximation (Fig.\,\ref{fig5}).

\subsection{BB model}

The two-band approximation in the BB model was considered in
\cite{Bobrov} using a free scaling factor $S$. Here we report the
results obtained with a fixed scaling factor ($S$=0.25) for both
directions (see Appendix B and Fig.\,\ref{fig20}). In contrast to
the GBTK approximation, the contribution $K$ of the smaller gap to
conductivity was not constant. Nevertheless the number of fitting
parameters was the same because at $T>$2~K $\gamma _{1}$ turns
zero in both directions and hence is no longer a fitting
parameter. Besides, unlike the broadening parameter $\Gamma $, the
pair-breaking parameter $\gamma $ does not increase the
uncertainty in calculating the r.m.s. deviation (cf.
Fig.\,\ref{fig18} with free $S$ and Fig.\,\ref{fig19}, Appendix).

\begin{figure}[htbp]
\includegraphics[width=8cm,angle=0]{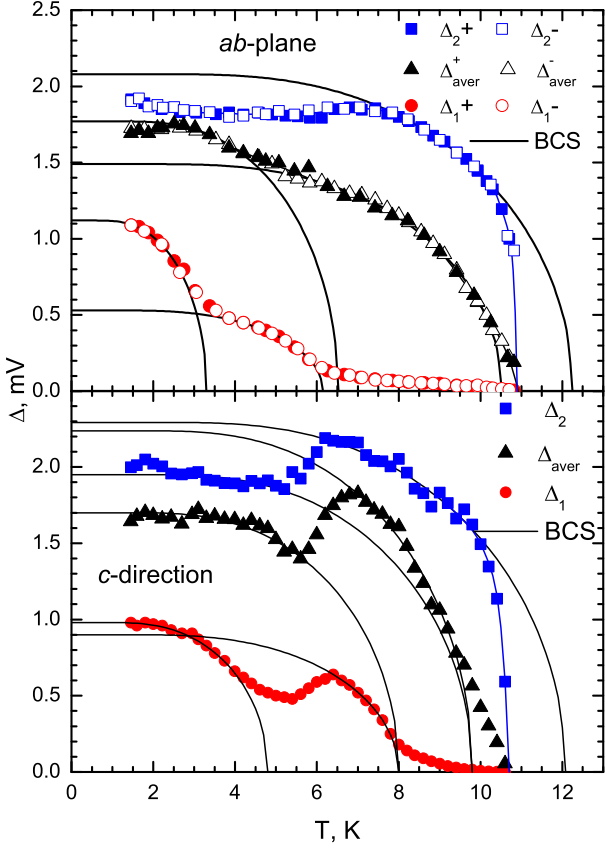}
\caption[]{Temperature dependencies of OPs $\Delta$  (BB model
\cite{Beloborodko}) calculated in the two-gap approximation by
Eq.(\ref{eq2}) for the PC in Fig.\,\ref{fig1}. The polynomial fit
is drawn through points outside the BCS approximation. $S$=0.25 in
both cases} \label{fig13}
\end{figure}

The temperature dependencies of the larger $\Delta _{2}$,  smaller
$\Delta_{1}$ and the average OP $\Delta_{\rm {aver}}$,
Eq.(\ref{eq3}), are shown in Fig.\,\ref{fig13}.

Note that the use of another model and a non-fixed $K$ to
conductivity affect the behavior of the temperature dependencies
of the OPs (mainly the smaller OP) in comparison to the GBTK model
and BB approximation with varying $S$ \cite{Bobrov}. In the
$ab$-plane the smaller OP decreases rapidly with temperature and
the BCS-extrapolation gives T$_{\rm c}\sim $3.3~K. At $T>$3~K
$\Delta _{1}$ in the AFM region changes into the BCS-dependence
with T$_{\rm c}\sim $6.15~K. In the PM region there is a region of
a smoothly decreasing gap which persists up to the normal state.
This region may be due to the interband interaction. In the
$c$-direction the BCS-extrapolated T$_{\rm c }$ values for $\Delta
_{1}$ are somewhat higher: $\sim $4.8~K (AFM region) and $\sim
$8~K (PM region).

The temperature dependencies of the larger OP $\Delta _{2}$
especially in the PM region closely resemble the shapes of the
curves for the OP calculated in the one-gap approximation within
the BB model with a varying scaling factor (Fig.\,\ref{fig9}). The
reason for the coincidence is quite obvious. In the PM region the
shape of the curve is mainly dependent on the larger OP because
its value is several times higher than that of the smaller OP. In
contrast to the one-gap approximation, the contribution of the
smaller OP to the conductivity holds the scaling factor $S$
constant. Since in the low temperature region OPs of different
bands are comparable in magnitude, the distinctions between the
larger OP of two-band calculations and the OP in the one-gap
approximation are more explicit in the shapes of the curves. The
critical temperatures obtained by BCS-extrapolation for $\Delta
_{2}$ in the PM region practically coincide with the one-gap
calculation (T$_{\rm c}\sim $12~K) and exceed the superconducting
transition temperature of the compound. On approaching $T\sim10$~K
the OPs start to depart from the BCS- dependence tending to 0 at
T$_{\rm c}$ of the sample. It is interesting that the temperature
dependencies of the average OP $\Delta _{\rm {aver}}$ have
sections in both directions in the PM region that decrease almost
linearly and go through zero at T$_{\rm c}$ of the sample.

\begin{figure}[htbp]
\includegraphics[width=8cm,angle=0]{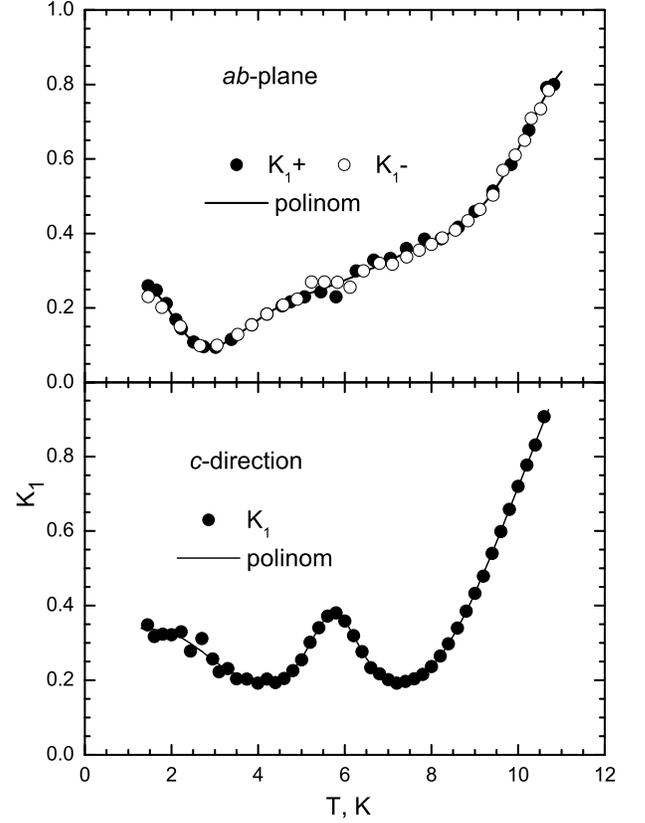}
\caption[]{Temperature dependencies of the contribution to
conductivity from the smaller OP (Fig.\,\ref{fig15}.) calculated
in the two-gap BB model} \label{fig14}
\end{figure}

Let us consider the temperature-dependent contribution $K$ to conductivity
made by the smaller gap (Fig.\,\ref{fig14} and Eq.(\ref{eq2})).

\begin{figure}[htbp]
\includegraphics[width=8cm,angle=0]{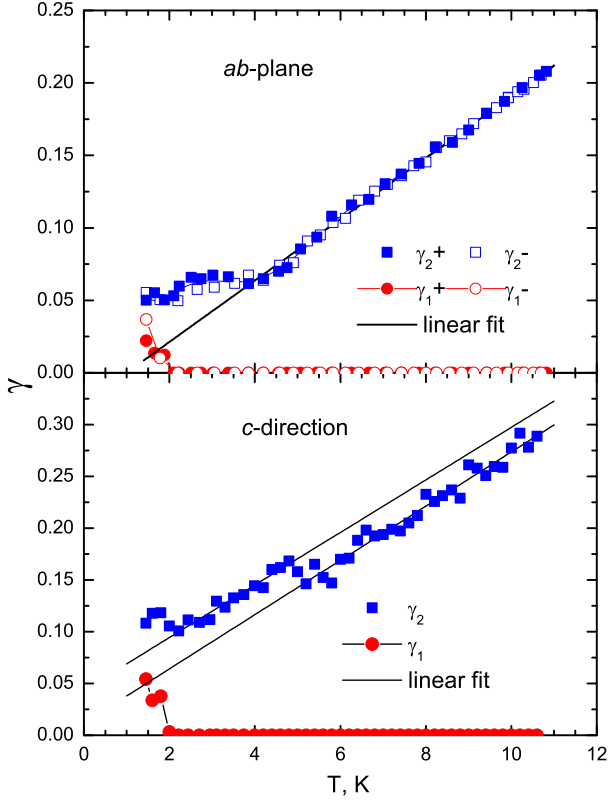}
\caption[]{Temperature dependencies of the pair-breaking parameter
$\gamma$ calculated in the two-gap BB model. $\gamma_{2}$ and
$\gamma_{1}$ correspond to the larger OP ($\Delta _{2}$) and to
the smaller OP ($\Delta _{1}$) respectively (Fig.\,\ref{fig13}).
For clarity, a linear fit is drawn through the points}
\label{fig15}
\end{figure}

At low temperatures the dependence $K(T)$ is readily predictable
qualitatively: $K$ decreases with a decrease in $\Delta _{1}$. On
a further rise of temperature these parameters exhibit a
correlated change in the $c$-direction: the growth of $\Delta
_{1}$ due to the reduced influence of spin density waves is
attended with an increase in $K$. The somewhat different degrees
of the changes in these parameters in the $c$-direction and the
absence of a similar correlation in the $ab$-plane might be
attributed to the increasing pair-breaking parameter for $\Delta
_{2}$ (Fig.\,\ref{fig15}) but this assumption is in conflict with
the sharp growth of $K$ in PM region. It is then, reasonable to
assume that $K$ grows because the relative share of the FS
containing the large gap decreases. This fact can be explained as
follows.

It is pointed above that the scaling factor $S$ is dependent in
particular on what part of the whole FS is occupied by
superconductivity. If $S$=const., this share is independent of
temperature. Also, the areas of the FS bands that contain a larger
and a smaller gap are also temperature-independent. This suggests
that if the FS area with a large OP decreases, superconductivity
in this part of the band is not suppressed fully. In terms of our
assumption, the OP in the ``vacant'' part of the band reduces and
becomes comparable to the OP in the second zone. We may thus
conclude that the superconducting share of the total FS area
remains invariable. This can account for the redistribution of the
relative FS shares between the large and small OPs. Near T$_{\rm
c}$ on approach of $T\sim10~K$ the contribution of the large OP to
conductivity falls below $\sim $20{\%}. As a result, $\Delta _{2}$
deviates from the BCS-dependence and turns rapidly to zero. The
physical reason for the reduction of the FS area with a large OP
may be connected with the spin fluctuations which enhance with
temperature. It is likely that larger OP occurs in the FS part
corresponding to the region in which the crystal lattice has
magnetic moments. This assumption is supported by the values of
pair-breaking parameters $\gamma _{2}$ of large OP which increase
with temperature. Note that the $\gamma _{2 {\-}}$ magnitudes of
larger OP calculated in the two-gap approximation within the BB
model for nonmagnetic LuNi$_{2}$B$_{2}$C are lower and decrease
(faster in the $ab$-plane) with temperature (see \cite{Bobrov1},
Fig.9). Since spin-flip scattering is absent in this compound, the
parameter $\gamma $ in this calculation tends to exceed the degree
of the superconducting gap broadening. Note that in
YNi$_{2}$B$_{2}$C the gap is most broadened in the low-energy part
(\cite{Nishimori}, Fig.6). Assuming that this reasoning holds for
ErNi$_{2}$B$_{2}$C too, we can conclude that no broadening of
larger and smaller gaps occurs above 2~K (Fig.\,\ref{fig15}) where
the parameter $\gamma $ is determined solely by the processes of
scattering at magnetic moments. It is therefore most justified to
apply the two-gap modification of the BB model in this case. The
temperature dependencies of broadening parameters $\gamma $ are
shown in Fig.\,\ref{fig15}. Two practically parallel parts of a
linear growth of $\gamma _{2}$ in the AFM and PM regions are
distinctly seen in the $c$-direction. There is only one linear
portion in the $ab$-plane, which may indicate that spin-density
waves are ineffective during a magnetic transition. Note that the
illustrated curves correlate to a certain degree with the curves
describing the contribution of a smaller OP to conductivity
(Fig.\,\ref{fig14}).
\begin{figure}[htbp]
\includegraphics[width=8cm,angle=0]{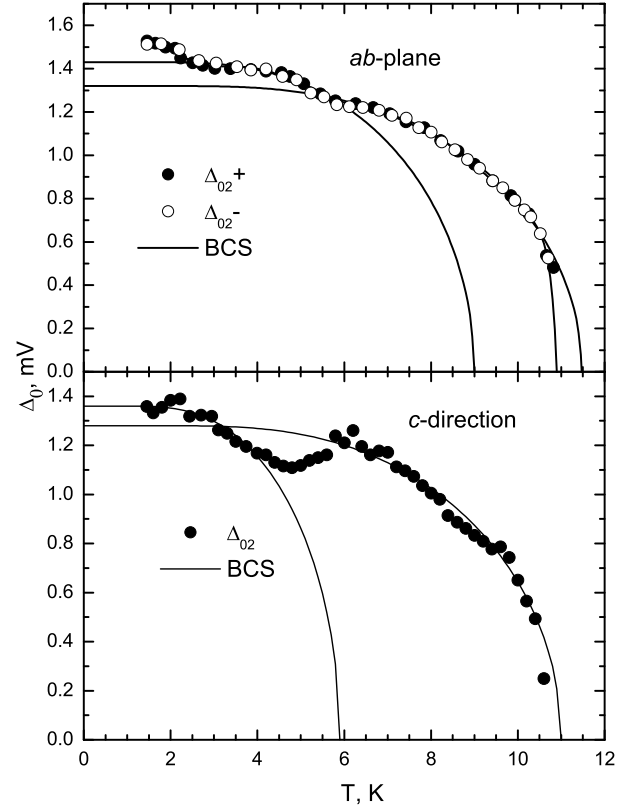}
\caption[]{The temperature dependence of the lager gap
 $\Delta _{02} =\Delta_2 \left(1-\gamma_2 ^{2/3} \right)^{3/2}$
(BB model) for the ErNi$_{2}$B$_{2}$C-Ag PC in the $ab$-plane and
the $c$-direction (Fig.\,\ref{fig13}). Solid lines are BCS
extrapolations. Since $\gamma_{1}$=0, the smaller gap behaves
similarly to the small OP.} \label{fig16}
\end{figure}

Fig.\,\ref{fig16} illustrates the temperature dependence of the
energy gap $ \Delta _{0}$ corresponding to $\Delta _{2}$ of the
larger OP. There are different BCS extrapolations for changing to
gapless superconductivity in the AFM (9~K and $\sim $6~K in the
$ab$-plane and $c$-direction, respectively) and PM ($\sim $11~K in
both directions) regions. The gap, like the OP, grows during the
transition to the PM state in the $c$-direction and has a
monotonic dependence in the $ab$-plane.

\begin{figure}[htbp]
\includegraphics[width=8cm,angle=0]{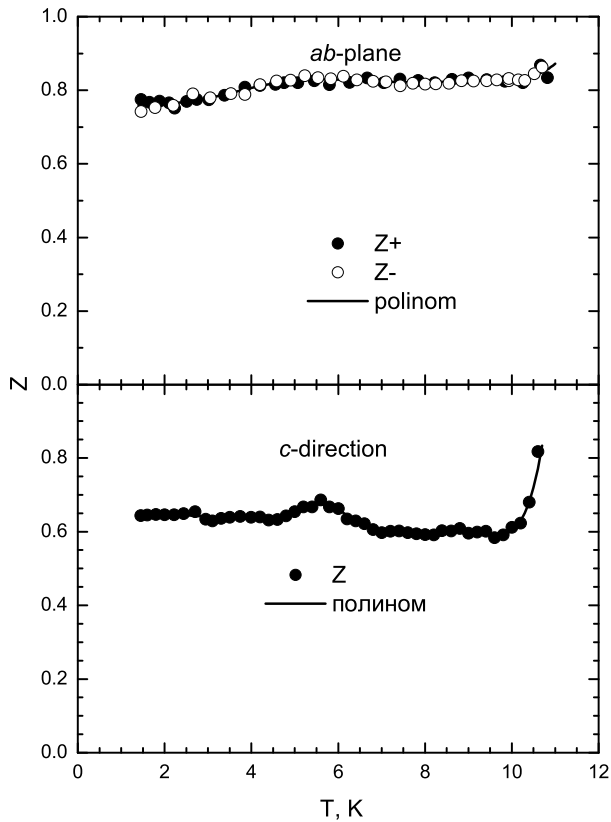}
\caption[]{Temperature dependencies of the tunnel parameter $Z$
(two-gap BB approximation for the ErNi$_{2}$B$_{2}$C-Ag PC in the
$ab$-plane and the $c$-direction).} \label{fig17}
\end{figure}

It is interesting that the tunnel parameter $Z$
(Fig.\,\ref{fig17}) has a feature at T$_{\rm N}$ only in the
$c$-direction, like in the one-gap case (Fig.\,\ref{fig9}), which
may be attributed to the effect of spin-density waves.

\section{Conclusions}

A detailed analysis of the temperature dependencies of PC Andreev
reflection spectra $dV/dI(V)$ has been performed for
ErNi$_{2}$B$_{2}$C (T$_{\rm c}\approx $11K) in the $ab$-plane and
$c$-direction using one-gap and two-gap approximations. Two models
were used: the traditional GBTK model including the broadening
parameter $\Gamma $ \cite{Plecenik} and the BB model
\cite{Beloborodko} in which the parameter $\gamma $ characterizes
the pair-breaking effect of magnetic moments (likely Er). For the
first time the calculation has been made comparing both the shape
and the intensity of experimental and theoretical curves. This has
decreased the degree of uncertainty in the temperature dependence
$\Delta $(T) for contacts with high broadening parameters $\Gamma
$.

The following conclusions have been drawn.

\begin{enumerate}
\item An anisotropic effect of AFM ordering has been detected
irrespective of the data processing model. For example, the
magnitude of the superconducting gap calculated within the one-gap
GBTK model decreases on transition to the AFM state by $\sim $
25{\%} in the $c$-direction and by $\sim $ 4-7{\%} in the
$ab$-plane, which correlates with the behavior of the averaged gap
in the two-gap approximation within the BB model \cite{Bobrov}.
\item The intensity of the PC spectra $dV/dI(V)$ changes in
correlation with the gap: the intensity decreases monotonically in
the $ab$-plane and has an extremum near the AFM transition in the
$c$-direction. This behavior may be due to the
orientation-dependent pair-breaking effect of spin-density waves.
Thus it has been found unambiguously that the AFM transition has
an anisotropic effect on the superconducting state. \item As in
\cite{Bobrov}, the pair-breaking parameter $\gamma $ increases in
the vicinity of magnetic transitions, which is natural to
attribute to the effect of spin fluctuations under a change of the
magnetic order. \item It has been shown that the proper choice of
the scaling factor $S$ in the one-gap GBTK calculation gives the
ratio 2$\Delta (0)$/kT$_{\rm c}\sim $3.52 for the gap in the PM
region and its BCS-like temperature dependence. The ratio obtained
in the one-gap calculation within the BB model is 2$\Delta
(0)$/kT$_{\rm c}\sim 4.08\div5.14$. \item The analysis of the
models shows that the two-gap calculation in the BB model with a
fixed scaling factor $S$ provides the most adequate information.
This calculation gives different BCS-extrapolation data for
T$_{\rm c}$ of the larger and smaller OPs, which points to the
multiband nature of superconductivity in ErNi$_{2}$B$_{2}$C. Its
physical sense is that there are FS parts with a weaker
electron-phonon interaction (EPI) in which the temperature induced
suppression of superconductivity is faster. This is supported by
the calculation of the anisotropy of the EPI parameter in
LuNi$_{2}$B$_{2}$C \cite{Bergk}, which can vary from 0.3$\div $0.8
on a spheroidal FS to 1.0$\div $2.7 on a cube-like FS.
\end{enumerate}

\section{Acknowledgement}
Work at Texas A\&M University was supported by the Robert A. Welch
Foundation, Houston Texas (Grant A-0514) and work in ILTPE was
supported by NAS of Ukraine.

\section{Appendix.}

\subsection{Some details of calculation technique}

The technique used (detailed in \cite{Bobrov1}) is based on a
selection of parameters that can ensure the smallest r.m.s.
divergence between experimental and theoretical curves. First, the
curves $dV/dI(V)$ were normalized to the curve $dV/dI(V)$ taken
above T$_{\rm c}$ and symmetrized. The curve fitting was performed
in the interval $\pm $8~mV to avoid the effect of the phonon
features (inflection points) in the vicinity of 10~mV in some
curves (see Fig.\,\ref{fig1}).

\subsection{Choice of proper scaling factor $S$}
Theoretically $S$=1. But $S<$1 happens very often too. The reasons
may be as follows.

\begin{enumerate}
\item The electron transport through a PC deviates from the
ballistic conditions \cite{Askerzade}. \item An inhomogeneity at
which the PC region at the S-electrode side is not fully
superconducting because of the normal region (regions) near the
$N-S$ boundary in the superconductor. \item A superconducting gap
can occur only in a part of the FS. For example, the AFM molecular
field induces a gap in some FS parts in ErNi$_{2}$B$_{2}$C and
suppresses the superconducting gap in these regions.
\end{enumerate}
\begin{figure}[htbp]
\includegraphics[width=8cm,angle=0]{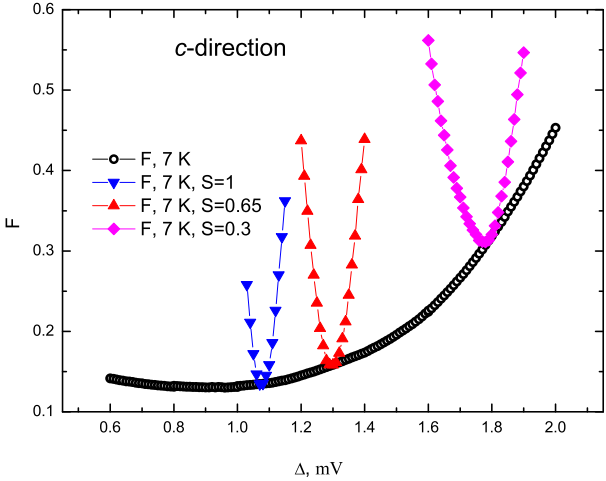}
\caption[]{The dependence of the r.m.s. deviation $F$ of the shape
of the theoretical curve from experimental $\Delta $ at 7\,K in
the $c$-direction. The intensity ratio $S$=($dV/dI(V))_{\rm
{exp}}$/($dV/dI(V))_{\rm {theor}}$ between experimental and
theoretical curves is changing from 3.58 to 0.23. The lowest $F$
corresponds to the best coincidence of the experimental and
theoretical curves is obtained at $\Delta $=0.96\,meV and $S$=1.3.
The narrow parabolic error curves were taken at $S$=const (one-gap
GBTK model \cite{Plecenik}} \label{fig18}
\end{figure}
A more exotic case of $S>$1 is also possible. For example, to
improve the description of the shapes of smeared experimental
curves, the parameter $\Gamma $ which assumes a finite lifetime of
carriers is usually increased. As a result, the intensity of
theoretical curves decreases. However, if the spectra $dV/dI(V)$
are smeared e.g., because of different values (distribution) of
the superconducting gap in different FS parts due to anisotropy or
multiband superconductivity, the theoretical $\Gamma $-broadening
curve can approach the shape of the experimental curve, but its
intensity will be lower, and the scaling factor $S$ can exceed 1.
This can be used as a criterion of validity for a model. As an
example, we consider the spectrum from Fig.\,14 in \cite{Bobrov2}.
The gap is distributed in the interval 1-3.35\,meV (insert). The
calculation in \cite{Bobrov1} gives:
\begin{enumerate}
\item GBTK (1-gap): $\Delta$=2.565~meV, $\Gamma $=0.523~meV,
$Z$=0.8, $S$=1.467. \item GBTK (2-gaps): $\Delta _{1}$=2.18~meV;
$\Delta _{2}$=2.99~meV; $\Gamma_{1}$=0.36~meV; $\Gamma
_{2}$=0.067~meV; $Z$=0.78; $K$=0.56; $\Delta_{\rm
{aver}}$=2.538~meV; $S$=1.144. \item BB (1-gap): $\Delta
$=2.79~meV, $\gamma $=0.04, $Z$=0.74, $S$=0.968. \item BB
(2-gaps): $\Delta _{1}$=2.15~meV; $\Delta _{2}$=2.984~meV;
$\gamma_{1}$=0.046; $\gamma _{2}$=0.011; $Z$=0.75; $K$=0.315;
$\Delta_{\rm {aver}}$=2.72~meV; $S$=0.968.
\end{enumerate}
Thus, the one-gap GBTK calculation results in the highest value in
the $S$ estimate. Therefore one-gap GBTK calculations \cite{Lu}
can be regarded as oversimplified and can be used only as a first
approximation. In ordinary superconductors, e.g., Zn
\cite{Naidyuk1}, $S\approx $1 is independent of temperature
because the gap opened isotropically on the whole FS at
$T<$T$_{\rm c}$.

In Fig.\,18 the smooth broad arc-like curve corresponds to the
best coincidence of the shapes of theoretical and experimental
curves for the gap $\Delta $ in the interval $0.6\div2$~meV. The
lowest error is at $\Delta $=0.96\,meV. The scaling factor varies
along the curve from $S=$3.58 at $\Delta $=0.6\,meV to $S$=0.23 at
$\Delta $=2\,meV and $S=1.3$ at the lowest $F$. $S>$1 is possible
only assuming the gap distribution (see above), and therefore the
\textit{lowest $F$ (shape error) alone is not sufficient to be a
criterion}. Besides, for the curves taken at higher $\Gamma$
values a comparatively small change in the shape of the
temperature-neighboring curves $dV/dI(V)$ can shift arbitrarily
the error minimum and the corresponding $\Delta$.

\begin{figure}[htbp]
\includegraphics[width=8cm,angle=0]{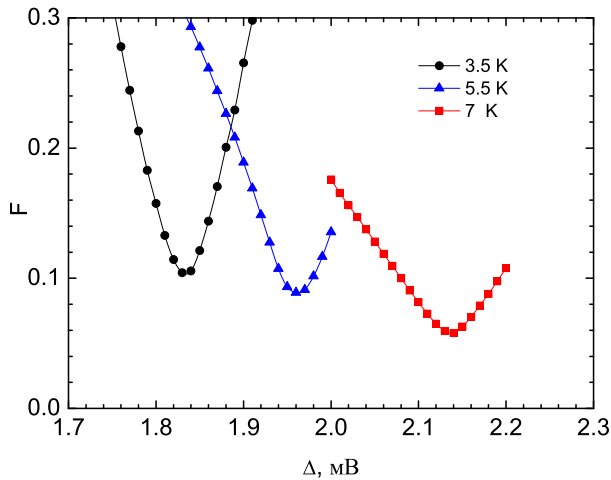}
\caption[]{The dependence of the mutual r.m.s. deviation $F$ of
the theoretical and experimental curve shapes from $\Delta $ (BB
model) in the $c$-direction at different temperatures. $S$ varies
along the curves and at the minima, $T$=3.5, 5.5 and 7~K, being
$S$=0.24, 0.2 and 0.22, respectively} \label{fig19}
\end{figure}
\begin{figure*}[htbp]
\includegraphics[width=0.75\linewidth]{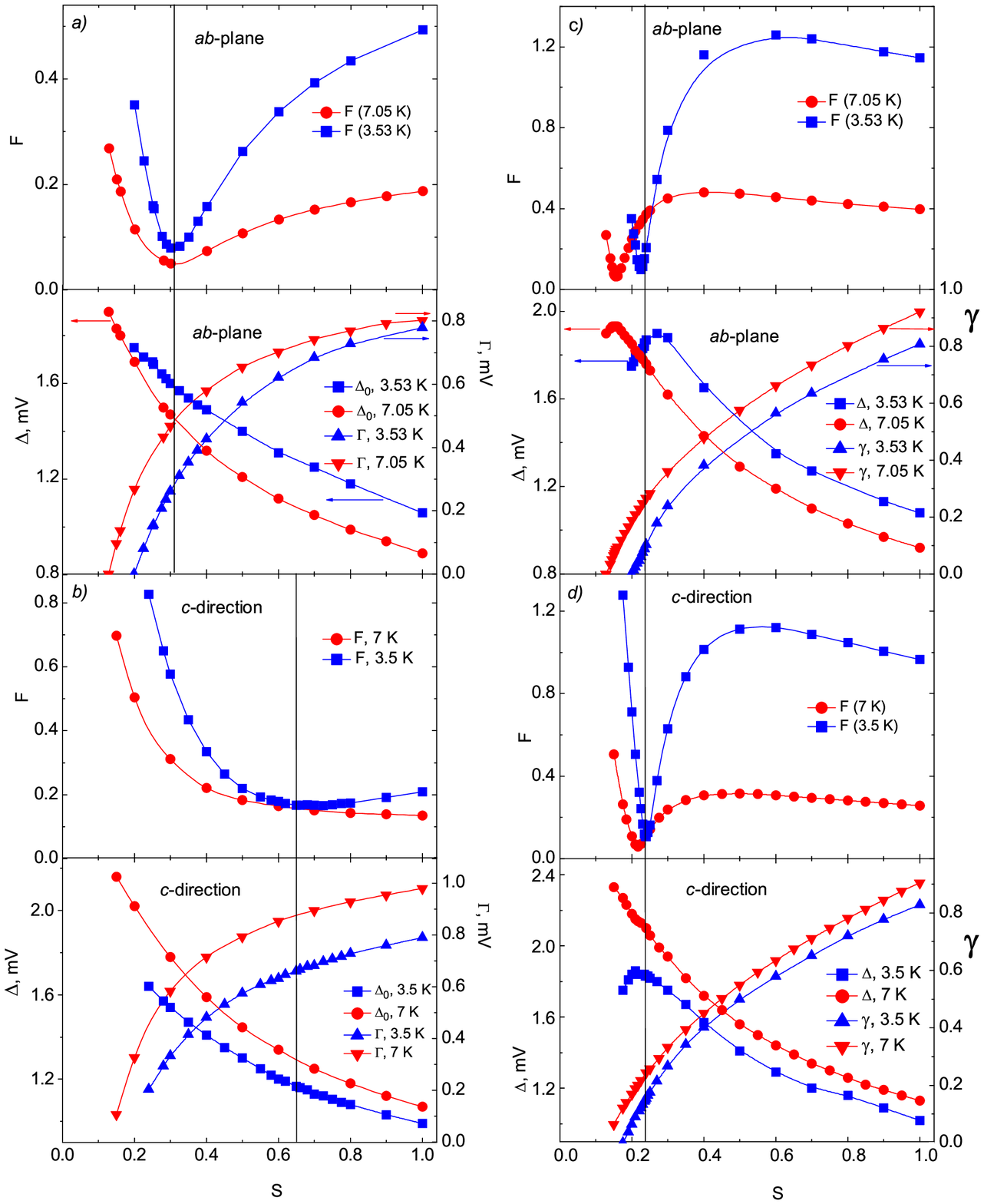}
\caption[]{Dependence of the relative deviation of the shapes $F$
of theoretical and experimental curves from $S$ calculated in the
GBTK (a, b) and BB (c, d) models at $T$=3.5~K and 7~K in the
$ab$-plane and in the $c$-direction. The lower part of each figure
shows the corresponding dependencies of the gap (OP) $\Delta $ and
the broadening $\Gamma $ (pair-breaking $\gamma )$ parameters. The
vertical line marks the best choice of $S$ corresponding to the
closest coincidence of the curve shapes at both temperatures}
\label{fig20}
\end{figure*}

Unlike the GBTK model, the lowest-error curves obtained in the BB
model have distinct minima (Fig.\,\ref{fig19}). In our case the
criterion of the proper choice of the scaling factor $S$ was its
value at which the shapes of the theoretical and experimental
curves coincided most closely in the whole interval of
temperatures used. To avoid overloading, Fig.\,20 contains the $F$
and $S$ calculation for two characteristic temperatures: $T$=3.5~K
in the middle of the AFM region and $T$=7~K, i.e. above the
temperature of AFM ordering. It follows from Fig.\,20 that $S$ can
vary from 1 to 0.2, being the lowest at $\Gamma \to $0
(Fig.\,\ref{fig20}).
The vertical lines show the best $S$-values corresponding to the
minimum error along with their associated gaps (OPs) and
broadening (pair-breaking) parameters. Note that in the BB model a
departure from the best-chosen $S$-value caused a sharp increase
in the error $F$, and therefore the limits of $S$ variation are
rather narrow. On the other hand, the GBTK model \cite{Plecenik}
allows more freedom of selecting $S$, especially in the
$c$-direction. However, this variation has practically no effect
on the shape of temperature dependence $\Delta (T)$ because the
curves $\Delta (S)$ are nearly parallel at both temperatures. The
gap can be estimated readily at another $S$ value using the
dependence $\Delta (S)$.

\subsection{One-gap BB calculation with S$\ne $\textit{const}}

Unlike the GBTK model, the lowest-error curves obtained in the BB
model have distinct minima (Fig.\,\ref{fig19}). Using a fixed $S$
impairs considerably the quality of fitting in this approximation.
The results of the one-gap calculation allowing for the minima in
the curves are illustrated in Fig.\,\ref{fig9} and the
dependencies $S(T)$ are shown in Fig.\,\ref{fig3}. The decrease in
the $S$-value is due to the two-gap character of superconductivity
in this compound, which determines different OP magnitudes and a
different (in comparison to the GBTK result) shape of the density
of states. The relative share of the FS with a larger OP decreases
at rising temperature (see the Two-Gap BB calculation section).
Used in BB theory the density of states terminates abruptly at the
gap edge. In the one-gap approximation of the experimental curve
this entails termination of the contribution to conductivity from
the FS parts with smaller OPs. However, since the spectrum
intensity is estimated assuming that the OP obtained is related to
the whole Fermi surface, $S$ decreases. The difference between the
dependencies $S(T)$ in the $c$-direction and the $ab$-plane can be
attribute to the anisotropic effect of spin-density waves (see the
above interpretation of the quantity $M$ and Fig.\,\ref{fig3}).
The temperature-dependent AFM molecular field induces a gap in
some parts of FS in the $c$-direction and destroys the
superconducting gap in them. An AFM gap is more probable in the FS
parts which have a smaller superconducting gap taking into account
the temperature dependence of the OP in the corresponding
temperature region.

The sharp decrease in the OP on approaching T$_{\rm c}$ can also
be caused by the two-band structure of the FS and the reduction of
the FS share with a larger OP. There may exist a certain minimal
FS share with a large OP below which superconductivity is
destroyed rapidly due to the interband interaction. In our case
the departure from the BCS-dependence started in both directions
at $T\sim $10\,K at which the contribution of the larger OP to
conductivity dropped below 20{\%} (two-gap calculation).

\subsection{Reducing the number of fitting parameters for the two-gap GBTK model}

As noted in the Introduction, in the general case the two-gap
calculation by Eq.(\ref{eq2}) involves seven fitting parameters
($\Delta _{1}$, $\Delta _{2}$, $\Gamma _{1}$, $\Gamma _{2}$, Z, S
and $K)$. This has an unfavorable effect on the estimates
obtained. It is therefore desirable to minimize the number of such
parameters on the basis of physically reasonable limitations. In
the strict sense, the scaling factor $S$ is not a fitting
parameter. It does not enter into the theoretical formulas
describing current-voltage characteristics. It is intended to
equalize the intensities of experimental and theoretical curves on
calculating the mutual r.m.s. deviation of their shapes. However,
a use of $S$=\textit{const} restricts the range of permissible
values for the rest of the fitting parameters. A fixed $S$ reduces
the number of fitting parameters at least by one parameter
(Fig.\,\ref{fig18}). We did not fix the tunnel parameter $Z$ which
was found to be approximately constant except near T$_{\rm c}$. As
a result, we have four fitting parameters instead of seven.
Besides, it is reasonable to fix the relative contribution $K$ to
conductivity from each band.

Thus, the calculation was made using fixed $S$=0.35 in the
$ab$-plane and $S$=0.5 in the $c$-direction. The relative
contribution of the smaller gap to conductivity was $K$=0.8 in
both directions. Fixed $K$ actually couples $\Delta $ and $\Gamma
$, which can distort their temperature dependencies. Indeed, if,
for example, the first gap remains constant and the second one
decreases, the relative contribution of the first gap increases
(it is assumed that the FS share of each gap does not change). To
exclude this, it is necessary to decrease the broadening parameter
of the second gap or to increase it for the first one. Identical
temperature dependencies for gaps and broadening parameters is the
simplest version free of distortions.

\section{Note added in proof}
After the paper was sent to arXiv, we were aware about recent
three-dimensional study of the Fermi surface of LuNi$_{2}$B$_{2}$C
in \cite{Dugdale}. This study shows that 1) the Fermi surface
topology of the rare-earth nickel borocarbides varies little for
rare-earth elements such as Er, Tm and Yb, 2) there are 3 bands
which contribution to the density-of-states (DoS) at the Fermi
energy is 0.24\%, 22.64\% and 77.1\%. That is, two bands basically
contribute to DoS and therefore our two-band approach is
reasonable.


\end{document}